\def\r#1#2#3#4#5{\hangindent 10 mm %
\noindent {\sc #1}, 19{#2}, {\it #3}, {\bf #4},
{#5}.}
\def\rr#1#2#3#4#5{\hangindent 10 mm %
\noindent {\sc #1}, 19{#2}, {\it #3}, {\bf #4}
{#5}.}
\def\newpar#1#2{\vskip5mm \begin{center}{\sc \S#1. #2}
\end{center} \nopagebreak}
\def\gesim{\lower 2pt \hbox{$\buildrel > \over \sim$}}
\def\lesim{\lower 2pt \hbox{$\buildrel < \over \sim$}}
\def\ea{{\it et al \/}}
\documentclass{article}
\begin{document}
%\baselineskip=1.8\baselineskip

%\pagestyle{headings}
\begin{center}
{\bf Peculiarities of phonon spectra and lattice heat capacity in Ir and Rh.}
\vskip 5mm

{\sc By M. I. Katsnelson$^\dag$, I. I. Naumov$^\ddag$, A. V. Trefilov,
M. N. Khlopkin and
K.~Yu.~Khromov} \vskip 5mm

{\footnotesize Russian Research Centre ``Kurchatov Institute'', Moscow 123182,
Russia \\
$^\dag$Institute of Metal Physics, Ekaterinburg 620219 ,Russia \\
$\ddag$Institue of Physics of Strength and Material Science,
Tomsk 644055, Russia } \vskip 5mm

{\sc Abstract}
\end{center}

{\footnotesize A simple pseudopotential model is proposed, which allows the
phonon spectra and temperature dependence of the lattice heat capacity of Ir
and Rh be described with a high enough accuracy. A careful comparison of the
calculated and experimental values of the lattice heat capacity
is carried out, with the procedure of the identification of the phonon
contribution to the heat capacity and determination of the characteristics
(momenta) of the phonon density of states from the experimental values of
the total heat capacity of metal at a constant pressure being described in
detail. The results of the theoretical calculations explain, in particular,
such peculiar feature of Ir and Rh, unusual for cubic metals, as a sharp
(more than by a factor of 1.5) decrease in the effective Debye temperature
with increasing termperature. The temperature dependence of the mean square
amplitude of atomic displacements in Ir and Rh has been calculated. Basing
on the band calculations the manifestation of the Kohn singularities in the
phonon spectra of Ir are discussed.}

\newpar {1} {Introduction}

From the physical point of view iridium and its analogue rhodium stand out
by their unusual mechanical properties among other FCC metals (see discussion
in Gornostyrev \ea 1994). The peculiarities of mechanical properties (at least
those inherent to high pure single crystals) should be eventually determined
by the specific features of interatomic interactions in metal. Greenberg \ea
1990 and Ivanov \ea 1994 succeded in constructing of a pseudopotential
model of pair forces for these metals wich describes surprisingly well
(for transition metals) their elastic properties. In terms of this model the
peculiarities of the Ir and Rh defect structure (Ivanov \ea 1994, Gornostyrev \ea
1994) could be analyzed, and the qualitative explanation of some of their
peculiar mechanical properties, primarily, the possibility of brittle failure
of single crystals after a long-term stage of plastic deformation
could be given. At the same time it was found that in the description of
the dispersion curves of phonons in Ir measured by Ivanov \ea 1994 the model
does not prove to be accurate enough for oscillation modes near the Brillouin
zone boundaries. According to the calculations of Ivanov \ea 1994 the
maximum frequency of the phonon spectrum appears to be overestimated by 25 \%
comparing with the experimental value found from the results of tunnel
experiments (Kraynukov \ea 1989), and flattenings in the
$\langle 110 \rangle$, $\langle 100 \rangle$ directions are observed, which
are not described by the proposed model.

The main purpose of the present paper is to ascertain the physical nature
of these descrepan\-cies, improve the model correspondingly, and calculate a
number of thermodynamical properties with the allowance for the corrected
phonon spectra.

It may seem that the use of a simple pseudopotential model for these purposes
does not correspond to the current state of the quantitative theory of
metals, when completely {\it ab initio \/} methods for calculation of phonon
spectra have been developed (see, e. g., review Maksimov and Savrasov 1995).
However, the description of a wide range of lattice properites, including
temperature dependence of different values is still beyond the capabilities
of completely {\it ab initio \/} approaches.
Moreover if it is taken into account that the pseudopotential approach
also gives the pairwise potentials which can be used for the simulation of defects,
diffusion processes, radiation effects, for the description of the liquid
phase properties (for example, the maximum value of the structural factor,
determined by the packing parameter) etc. then the importance of the
development of the pseudopotential theory for the metals for which it is
valid appears indisputable. An important advantage of this approach is
also its simplicity enabling many complicated problems to be discussed
not only at the qualitative but also at the quantitative level, and the
close relationship with the chemical bond language used in the material
science (Gornostyrev \ea 1994). For the clarification of the physical
mechanism of the Ir dispersion curve flattening the results of band
calculations are given in \S2, on which basis the Kohn singularities in the
phonon spectra are discussed. In \S3 a modification of the pseudopotential
form permitting these effects to be partially simulated for the purpose of
an optimal description of the phonon spectra is proposed. In \S4 the
characteristics of the phonon spectra extracted from the experimental data
on the Ir and Rh heat capacity are given with a detailed description
of the corresponding treatment procedure. \S5 presents the calculation
results of the lattice heat capacity, and their comparison with the
experiment. \S6 summarizes the results of the paper.

\newpar {2} {Fermi surface and Kohn singularities in Ir.}

Generally, one could think that the high accuracy of the description of
elastic moduli, and hence, of phonon spectra at small wave vectors $q$,
provide a good agreement of the theory with experiment for the dispersion
curves in the larger part of the Brillouin zone. Significant deviations from
the ''smooth'' behavior of the phonon frequencies depending on the
wave vector {\bf q} can be connected, first of all, with the Kohn
singularities (Izymov and Chernoplekov 1995) when $q$ coincides with the
Fermi surface diameter $2k_F$ in the corresponding direction. However in the
three dimensional case the typical Kohn singularities in the phonon
frequencies $\omega({\bf q})$ are of the order $(q-2k_F) \ln |q-2k_F|$
and are weakly expressed. The Kohn singularities clearly expressed in
$\omega({\bf q})$ can be only observed in the case of the Fermi surface
flattening in one (cylindrical sections) or two (flat sections)
directions when
$\delta\omega({\bf q})\sim \pm \sqrt{|q-2k_F|} \Theta (\pm (q-2k_F))$
($\Theta(x>0)=1, \Theta(x<0)=0$)  or
$\delta \omega({\bf q})\sim\ln |q-2k_F|$,
respectively (Katsnelson, Naumov and Trefilov 1994). Figs. 1,2 present
the results of our calculations of the Fermi surface of Ir and nonuniform
susceptibility
$$ \chi ({\bf q}) =2 \sum_{nn'} \int {d {\bf k} \over (2\pi)^3}
{f (\varepsilon_n ({\bf k})) [1-f (\varepsilon_{n'} ({\bf k}+{\bf q}))] \over
\varepsilon_{n'} ({\bf k}+{\bf q}) - \varepsilon_n ({\bf k}) }
\eqno (1) $$
where $\varepsilon_n ({\bf k})$ is the electron energy spectrum; $n$, $n'$
are the band indices; $f(\varepsilon)$ is the Fermi function; integration
over {\bf k} goes over the Brillouin zone (calculation method see in review
by Katsnelson \ea 1994). The function $\chi({\bf q})$ determines the form and magnitude
of the Kohn singularities in $\omega({\bf q})$. In fig. 1 the flattened
sections of the Fermi surface in the directions $\langle 100 \rangle$
and $\langle 110 \rangle$ are clearly seen. Fig. 2 shows the maxima of
$\chi ({\bf q})$ at the relevant values of wave vectors (in the case of the
spherical Fermi surface the maxima should have manifested themselves only
in $d \chi / dq$). At the same time in the calculations of the phonon spectra
in terms of the pseudopotential model in the second order of the perturbation
theory (Greenberg \ea 1990, Ivanov \ea 1994) the character of the  Kohn
singularities corresponds to the assumption about the sphericity of the
Fermi surface and, therefore, the phonon spectrum in the relevant region
of the {\bf q} space, naturally, proves to be too rigid. If we want to improve
the description of phonon spectra in terms of the simple pseudopotential
approach, the softening of the phonon frequencies due to the non-sphericity
of the Fermi surface can be simulated and described by means of changing
the form of the Fourier image of pseudopotential $V_{ps}(q)$
at $0.6 (2\pi /a) \lesim q \lesim (2\pi /a)$ ($a$ is the lattice constant),
determined from the band calculations results. It is this circumstance
that should be primarily taken into account in changing the pseudopotential
model for Ir and Rh.

\newpar {3} {Calculation of Ir and Rh phonon spectra}

In the calculation of phonon spectra, elastic moduli, and the interatomic
interaction potential in papers by Greenberg \ea 1990, Ivanov \ea 1994
one of the common pseudopotential models (Heine and Weaire 1970; Browman
and Kagan 1974; Vaks and Trefilov 1977)
was used. The Fourier component
of the local pseudopotential of the electron-ion interaction was taken in the
form of the Animalu-Heine pseudopotential
$$ V^{AH}_{ps} =- {4 \pi Z \over q^2 \Omega_0} \left [ \cos q r_0 -V_0
\left ( {\sin q r_0 \over q r_0 } -\cos q r_0 \right ) \right ]
\exp \left [ -\xi \left ({q \over 2 k_{F0}} \right )^4 \right ] \eqno (2)
$$
where $\Omega_0$ is the atomic volume, $Z$ is the effective ion charge,
$r_0$, $V_0$ are the fitting parameters determined from the condition of
the agreement with experiment $\Omega_0$ and shear modulus $B_{44}$
at pressure $p=0$; $\xi$ is the constant whose value is selected
proceeding from the consideration of fast enough convergence of the sums
over the reciprocal lattice vectors (here and below the atomic units are
used). The screening function of the electron gas was taken in the
approximation by Geldart and Taylor 1970. See Greenberg \ea 1990, Ivanov \ea
1994 for the details of the theoretical model, including the substantiation
of the choice of values $Z=4.5$ for Ir and $Z=3.86$ for Rh. As discussed in
the introduction this model provides a high enough accuracy of the
description of elastic moduli and much lower one in the description of the
phonon frequencies near the Brillouin zone boundaries. The latter is the
result of neglecting the nonsphericity of the Fermi surface, which gives
no possibility to describe adequately the Kohn singularities in the
phonon spectra (see \S\ 2).

For the purpose of simulating these effects remaining in the framework of
the simplest pseudopotential model, we considered various possible changes
in the form of the psedopotential form factor
$$ V_{ps} (q) =V_{ps}^{AH} (q) +\delta V_{ps} (q)  \eqno (3) $$
at large $q$. It is seen
from the results of \S2 that the ``dangerous'' region of the   wave vectors,
containing the strongest Kohn singularities (which is the result of the
maximum non-sphericity of the Fermi  surface in the relevant directions)
is approximately $(0.5 \le g_1 \le 0.7) g_1$, where $g_1$ is the
minimum vector
of the reciprocal lattice. The addition to the pseudopotential
$\delta V_{ps} (q)$ to simulate these singularities should be noticable
in the relevant region of $q$ values and vanishingly small at
$q=0$ and $q \ge g_1$. It is natural to expect that the addition
$\delta V_{ps} (q)$ would not noticably affect the calculation values of
pressure and elastic moduli which are solely determined by the behavior
of $V_{ps} (q)$ at $q \to 0$ and by $V_{ps} ({\bf g })$ for the reciprocal
lattice vectors. The concrete form of the addition $\delta V_{ps} (q)$
is rather arbitrary, however it should affect the $V_{ps} ({\bf q })$
in the region of $q$ specified at the end of \S\ 2.
The following dependences satisfying the above mentioned condition
were actually studied:
$$\delta_1 V_{ps} (q) = \beta (qr_0) \exp [-\alpha (qr_0)^4] \eqno (4a) $$
$$\delta_2 V_{ps} (q) = \beta (qr_0)^2 \exp [-\alpha (qr_0)^4] \eqno (4b) $$
%
%$$ \delta_3 V_{ps} (q) = \beta \exp [-\alpha (qr_0)^4] \eqno (4c) $$
%
$$ \delta_4 V_{ps} (q) = \beta (qr_0)^2 \exp [-\alpha (q-q_0)^2 r_0^2] \eqno
(4c) $$
$$ \delta_5 V_{ps} (q) = \beta (qr_0)^2 \exp [-\alpha (q-q_0)^4 r_0^4] \eqno
(4d) $$
where parameters $\beta$, $\alpha$ (as well as $r_0$, $V_0$) are found from
conditions
$\Omega_0= \Omega^{exp}_0$, $\Theta_{LT}=\Theta_{LT}^{exp}$, $\omega_X=
\omega_{max}^{exp}$. However, exact fitting to these conditions turned out
to be impossible, so the set of parameters giving the best approximation
was used.
Quantity $\omega_X$ is the frequency of
phonons at the $X$ point of the Brillouin zone
(this is the point where the phonon spectrum calculated
using the Animalu-Heine pseudopetential used as a basis for the
fitting procedure reaches its maximum). The  experimental values
of $\omega_{max}$ were taken from results of the tunnel
experiments by Kraynukov \ea 1989 for Ir and Zhalko-Titarenko \ea 1989
for Rh. The experimental of $\omega_{max}$
and calculated  values of $\omega_X$
are listed if table 2 for Ir  and table 3 for Rh. $q_0$ is the wave
vector at the point X of the Brillouin zone. Among the additions
(4a)--(4d) (4c) gives the best agreement of the calculated phonon spectra
with the experiment, although the choice of the addition if the form (4a)
permits nearly the same agreement between the theory and the experiment
to be reached. It is important to emphasize that parameter $\beta$
is choosen so that additions (4) were small without giving significant
distorsion of the $V_{ps}(q)$ form when comparing with the standart
Animalu-Heine model. Nevertheless, calculation results for the phonon
spectrum (but not for $q \to 0$, i. e. not for elastic moduli) proves to be
very sensitive to the explicit form of these small additions. The sensitivity
of the calculation results to the form of the pseudopotential will be
considered in detail in a separate paper. Table 1 lists the parameters
for two versions of the pseudepotential,
$V_{ps}^{AH}(q)$ and $V_{ps}^{AH}(q)+\delta V_{ps}(q)$, where addition
$\delta V_{ps}(q)$ is choosen as (4c) which, comparing with the rest
possibilities investigated gives the best description of the lattice
properties of Ir and Rh. Tables 2 and 3 and figs. 3,4 give the results
of the corresponding calculations. $\Theta_{LT}=\Theta_D(T=0)$
is the low temparature value of the Debye temperature calculated from
the theoretical values of $B_{ik}$, using as the bulk modulus $B$ its static
value $B_{st}$ determined by differentiating the total energy over volume.
Quantity $B_{dyn}$ obtained from the phonon spectra by extrapolation has a
methodical meaning: difference $|B_{st}-B_{dyn}|$ characterizes the value
of multiion interactions (Browman and Kagan 1974). For Ir and Rh its value
is substantially smaller than for other transition metals (Vaks \ea 1989).
Quantity
$\gamma_0= -\partial \ln \Theta_{LT} / \partial \ln \Omega$ is the low
temperature Gruneisen parameter. Note the high sensitivity of $\gamma_0$
to the form of the pseudopotential (see tables 2--3). We will discuss
the sensitivity of results to the form of the pseudopotential in a
separate paper. Unfortunately, at present we don't have experimental results
to compare with the calculated values of $\gamma_0$.
The arrows in figs. 3,4 show the positions
of the Kohn singularities in the approximation of free electrons
$|{\bf q}+{\bf g}|=2k_F$. An unusual feature of the considered phonon
spectra is that in one of the branches in the X-W directions
$\omega({\bf q})$ is practically independent on $q$. This feature is not
connected with the Kohn singularities in any way and is obtained even in
the approximation of the nearest neighbours for interatomic interactions.
As discussed in papers by Ivanov \ea 1994, Gornostyrev \ea 1994 this approximation
describes
reasonably the elastic moduli of Ir.

It is seen that by introducing ``simulation'' corrections into $V_{ps}(q)$
one can reach a better description of the phonon spectra of Ir and
apparently Rh. This allows us to expect that the calculated
phonon spectra can be used for obtaining true enough information on the
thermodynamical properties of these metals. It should be emphasized that it
is difficult to obtain the information about the phonon spectra of Ir using
the method of inelastic neutron scattering because of the large cross section
of neutron absorption. Therefore when considering the agreement between the
theory and experiment, it is important to get the information on the Ir
lattice dynamics from the analysis of the temparature dependences of the
thermodynamical properties of Ir, primarily its heat capacity. In the next
section we shall use this procedure for obtaining the values of different
characteristics (momenta) of the Ir and Rh phonon spectra.
\newpage

\newpar{4} {Analysis of experimental data on Ir and Rh heat capacity}

The lattice heat capacity is determined by the form of phonon density
of states $g(\omega)$. Some characteristics of the latter, which can be used
for the comparison of our description of phonon spectra with the experiment,
can be derived from the analysis of the temperature dependence of the heat
capacity at a constant pressure $C_p(T)$.

In this section we identified the lattice component in the heat capacity
measured in the experiment and determined values of some average
frequencies (momenta) of the phonon spectra.

We used the following set of data in the analysis: in region 12--250 K ---
direct experimental data (Clusius and Losa 1959) and in region 1.2--12 K ---
the data from the handbook by Hultgren \ea 1973, using the data by Wolcott
1956 for the region 1.2--20 K.

In the analysis we neglected the temperature dependence of the coefficient
of the electron heat capacity and assumed that the anharmonicity is weak.
The assumption on the smallness of the anharmonicity contributions to the
thermodynamical properties of metals is justified as it was verified by
direct calculations and proved to be true even for such metals with soft
phonon modes as alkali ones (Vaks, Kravchuk and Trefilov 1980). With these
assumptions the experimentally measured heat capacity at constant pressure
$C_p$ is described by the relationships
$$
C_p = C_{ph} + C_a   \eqno(5a)
$$
$$
C_a=\left\{\gamma + ( A - \gamma)(C_{ph}/3R)^2\right\} T \eqno(5b)
$$
where $ C_{p}$ is the molar heat capacity at constant pressure;

$ C_{ph}$ is the phonon component of the heat capacity in the harmonic
approximation;

$C_a$ --- contains the contributions to the heat capacity, having the linear
temperature depen\-dence caused by anharmonic effects, thermal expansion of
the lattice as well as by the conduction electrons; and here

$\gamma$ is the coefficient of the electron heat capacity at low temperatures;

$A$ is the coefficient at the linear temparature term at high  termperatures;

$R$ is the gas constant.

Interpolation formula (5b) for $C_a$ gives the relevant linear temperature
asymptotes both at low and high temperatures and ensures a smooth transition
between the low- and high temperature asymptotes by a law similar to the
Nernst-Lindeman formula (Reznitskii 1981).

In the high temparature region the phonon component of the heat capacity
was described by the expression proposed by Naumov (Naumov 1994):
 $$
 C_{ph}=3R\left\{1-\frac{1}{12}\left(\frac{\Omega_2}{T}\right)^2+
\frac{1}{240}\left(\frac{\Omega_4}{T}\right)^4 +
\varphi\left(\frac{\Omega_*}{T}\right)\right\} \eqno (6)
$$
$$
\varphi(z)=\frac{z^2exp(z)}{\left(1-exp(z)\right)^2}-
\left(1-\frac{1}{12}z^2+ \frac{1}{240}z^4\right) \eqno (7)
$$
Here the asymptotic expansion of the phonon heat capacity by small
parameter $z=\Omega/T$ is used:
 $$
 C_{ph}=3R\left(1-\sum_{n=2}^{\infty}\frac{(n-1)B_n}{n!}\left
(\frac{\Omega_{n}}{T}\right)^{n}\right) \eqno (8)
 $$
where $B_{n}$ are the Bernulli numbers ($B_2=1/6$, $B_4=-1/30$,
$B_6=1/42$, $B_8=-1/30$, $B_{10}=5/66$
and for all odd $n$ all $B_n=0$ beginning from $n=3$).

In (6) corrections of second and fourth order by $\Omega/T$ are written
separetaly, and function $\varphi (\frac{\Omega_*}{T})$ allows for all higher
order corrections in the ``Einstein'' approximation, i. e. setting
$\Omega_{n} = \Omega_{*}$ for all $n \ge 6$.

Quantities $ \Omega _{n} $ characterize the momenta (average frequencies)
of the phonon spectrum according to the relationship
 $$
 (\Omega_{n})^{n}=<\omega^n>=\left.\int_0^{\infty}g(\omega)\,
\omega^{n}\,d{\omega}\right/\int_0^{\infty}g(\omega)\,d{\omega}, \eqno (9)
 $$
$g(\omega)$ being the phonon density of states.

Quantity $\gamma$ in $C_a$, (5b) was determined by the standart procedure:
by the approximation of the heat capacity in the low temparature region
(particulary 1.2--15 K) by a relationship such as
$C_p=\gamma T+\beta T^3+\alpha T^5$. Quantities
$ \gamma$, $\beta$ and $\alpha$ as well as the value of the Debye temperature
in the low temperature limit $\Theta_{LT}$, related to $\beta$ by the
equality $\beta=12\pi^4R/(5\Theta^3_{LT})$ are listed in the table 4.
The estimates of $\gamma$ and $\Theta_{LT}$ obtained from the set of all used
data (Clusius and Losa 1959, Wolcott 1956) agree resonably well with the
data available from the literature (see table 4).

As for quantities $A$, $\Omega_2, \Omega_4$ and $\Omega_*$ they were
determined by the method of least squares, approximating the heat capacity
by relations (6,7) in the temperature region 50--250 K. Within this
temperature range these relationships described the experimental results of
Clusius and Losa 1959 with a mean-square deviation of about 0.6 \%. The values
of parameters
$A$, $\Omega_2$,
$\Omega_4$ and $\Omega_*$, determined by the least square method are listed
in table 4. The table also gives the value of the Debye temperature in high
temperature limit $\Theta_{HT}=\Theta_D(T)$, where
$\Theta_{LT} \le T < T_m$ ($T_m$ is the melting temperature) related to the
second momentum of the phonon spectrum by:
$\Omega_2 = \Theta_{HT} \sqrt{3/5}$ (Maradudin, Montroll and Weiss 1963).

The described analisys permitted phonon contribution $C_{ph}$
in the harmonic approximation
to be
separated from the total heat capacity in the temperature region below
273 K, excluding the electron and anharmonic contributions. Some momenta
of the phonon spectra are expressed directly by the integrals of the phonon
heat capacity (Junod 1980).
 $$
<\omega>=2\int_0^{\infty}\left(1-\frac{C_{ph}}{3R}\right)\,dT
$$
$$
<\omega^{-1}>=\frac{3}{\pi^2}\int_0^{\infty}\frac{C_{ph}}{3R}\,T^{-2}\,dT
$$
$$
<\omega^{-2}>=0.138651\int_0^{\infty}\frac{C_{ph}}{3R}\,T^{-3}\,dT  \eqno(10)
$$
$$
<\omega^{-1}\,log\,\omega>=\frac{3}{\pi^2}\int_0^{\infty}\frac{C_{ph}}{3R}\,log\left(\frac{T}{0.70702}\right)\,T^{-2}\,dT
$$

We have calculated these momenta. In the region 15-273 K the integration was
carried out by the experimental points (Clusius and Losa 1959) while outside
this region we extrapolated the heat capacities by a low temperature
asymptote $C_{ph}=\beta T^3+\alpha T^5$ and high temperature asymptote in the
model of the Debye spectrum as in Mirmelstein \ea 1984. The average
frequencies corresponding to this momenta are listed in table 4. The
meaning of the quantities $\Omega_{-2}$, $\Omega_{-1}$,  $\Omega_1$
corresponds to the meaning of $\Omega_n$ in relationship (9) for
$n=-2,-1$ and $1$, while the meaning of $\Omega_{log}$ is determined by
 $$
\log(\Omega_{log})=\frac{<\omega^{-1}\,\log\omega>}
{<\omega^{-1}>}=\left.\int_0^{\infty}\frac{g(\omega)
\log{\omega}}{\omega}\,d{\omega}\right/ \int_0^{\infty}
\frac{g(\omega)}{\omega}\,d{\omega}
$$

\newpar {5} {Peculiarities of the temperature dependence of heat capacity of
Ir and Rh}

Although the directly measured quantity is heat capacity, it is more
convinient to discuss its temperature dependence via the temperature
dependence of the effective Debye temperature $\Theta (T)$. This is related
to the circumstance that the $C_{ph} (T)$ reaches the asymptotic $3R$ rather
quickly with the increase of $T$ (and this result will be reproduced in any,
even apparently incorrect model of the phonon spectrum, and the dependence
$\Theta (T)$ is a finer characteristic determining, in particular, the rate
at which $C_{ph} (T)$ reaches the asymptotic (Maradudin, \ea
1963).

An unusual feature of Ir and Rh, comparing with the other FCC metals, is the
sharp fall of the effective Debye temperature with temperature
($\Theta_{LT} / \Theta_{HT} \gesim 1.5$). For example according to handbook
by Zinoviev 1989 for Ca
$\Theta_{LT} \approx \Theta_{HT} \approx 230\, K$, for Cu
$\Theta_{LT} \approx 342 \,K$, $\Theta_{HT} \approx 310\, K$, for Pd
$\Theta_{LT} \approx 283 \,K$, $\Theta_{HT} \approx 275\, K$. Only in Ni
ratio
$\Theta_{LT} / \Theta_{HT} \approx 476\,K / 345\, K \approx 1.4$
differs strongly enough from unity (note, by the way, that in this handbook
Ir has a very underestimated value of $\Theta_{HT}$).

For comparing  the experimental data on $C_{ph} (T)$ in Ir and Rh the
calculations were made using phonon spectra $\omega_{{\bf q} \nu}$
that we had obtained
$$ C_{ph} (T) =R \sum_{{\bf q} \nu} \left ( {\hbar \omega_{{\bf q} \nu}
\over T} \right )^2 {\exp ( \hbar \omega_{{\bf q} \nu}  /T ) \over
[\exp ( \hbar \omega_{{\bf q} \nu}  /T ) -1 ]^2 } \eqno (11) $$
where the summation over the wave vector is made over the Brillouin zone
and $\nu$ is the number of the branch. The effective Debye temparature was derived
from the standart equation (Maradudin, \ea
1963).
$$C_{ph} (T)=9R \left [ {T \over \Theta(T)} \right ]^3 \int\limits_0
^{\Theta(T)/T} {dx x^4 e^x \over (e^x-1)^2} \eqno (12) $$
The calculation results are shown in figs. 5,6 and in table 4.
It should be
noted  that  the corrected pseudopotential with the addition of the
form (4c) describes average frequencies $\Omega_n$ much better than the
Animalu-Heine pseudopotential proposed by Ivanov \ea 1994. For example,
the agreement between the theoretical and experimental values of
$\Omega_1$  for Ir improved from 17 \% to 9\% and similarly for the rest
$\Omega_n$. It should be emphasized that in the calculation of $C_V(T)$
the change in the volume  with temperature was not taken into account.
It may be assumed that taking into account this dependence could improve
the agreement between the theory and experiment by some more percent.
The difference between the experimental and theoretical values of effective
Debye temperature
($|(\Theta_{exp}-\Theta_{theor})/
\Theta_{theor}|$) does not exceed 10 \%. It seems surprising that a simple
pseudopotential model can provide the description of the lattice heat
capacity of transition metals (Ir and Rh) at an accuracy level comparable
with such for alkali metals. The fact is really unusual as it is commonly
believed that the theory of pseudopotential in its simplest form used in
this paper is a priori not applicable to transition metals and has not so
high accuracy even for polyvalent $s-p$ metals (e. g. Al). Fig 7 shows the
corresponding data on $\Theta_D(T)$ for Na obtained from the calculation
results (Vaks \ea 1978). The accuracy of the description of the phonon
spectra in the pseudopotential model for other transition metals is
incomparably lower while, say, for Ni this model is totally unsuitable
(Vals \ea 1989).

When phonon spectrum $\omega_{{\bf q} \nu}$ is known in the whole Brillouin
zone one can calculate the temperature dependence of the average square
of the atomic displacements in the harmonic approximation
$$ \overline {x^2(T)} = \sum_{{\bf q} \nu} {\hbar \over 2M
\omega_{{\bf q} \nu}} \mbox{cotanh} \left ( {\hbar \omega_{{\bf q} \nu}
\over 2T} \right )\eqno (13) $$
where $M$ is the  atom mass. The corresponding results are  shown in fig. 8.
As far as we know the experimental data on $\overline{x^2(T)}$ in Ir and Rh
are currently not available.

\newpar {6} {Conclusion}

In conclusion we shall underline the most important results obtained in the
paper. The most surprising and interesting finding appears to be that
the simplest physical pseudopotential model, we proposed earlier, with
appropriate modifications seems to provide a high accuracy of the description
of phonon spectra and lattice heat capacity of transition metals Ir and Rh
and the accuracy of description is comparable to that for alkali metals.
This confirms our   earlier consideration (Greenberg  \ea 1990, Gornostyrev
\ea 1994) about the applicability of the model of pair interactions for the
description of the lattice properties of Ir and Rh. However the effects of
non-sphericity of the Fermi surface, substantial for all transition
metals cannot be completely accounted for. Nevertheless in Ir (and, as
appears in Rh) the role of these effects is reduced to some distorsions of
the despersion curves in directions in the Brillouin zone
($\langle 100 \rangle$  and $\langle 110 \rangle$) and appears to be
insignificant in the integral characteristics, i. e. in the heat capacity.
This situation would be absolutely impossible if in these metals  the Fermi
level were at the peak of the density of states $E_F$, as, say in
Ni or Pd, because
this peak affects substantially the elastic moduli and phonon
frequencies in the noticable part of the Brillouin zone (Katsnelson \ea
1994). The band calculations show, however (Greenberg \ea 1990) that not
only peak of $N(E)$ does not exist   but also there are no
relatively weak van Hove singularities near $E_F$ in Ir. Of course in itself
it cannot be taken as a direct evidence of the applicability of the
psudopotential model (in W, say, $N(E)$ near $E_F$ is also a smooth function
but a reasonable model pseudopotential cannot be constructed). However this
can be considered as a some indication to the possibility of such simple
description. In the whole, to solve the question of its applicability a wide
range of the lattice properties (thermal expansion, equation of state at
finite temperatures, other anharmonic effects, etc.) has to be studied.
We shall consider these questions in separate papers.

The study described in this paper became possible partially due to the
financial support of the International Science Foundation (Grant RGQ300)
and Russian Fund for Fundamental Research (Grant 95-02-06426).

\newpage
\begin{center} {\sc References} \end{center}

%1
\r{Browman, E. G., and Kagan, Yu. M.} {74} {Uspekhi Fiz. Nauk} {112} {369}

%2
\r{Budworth, D. W., Hear F. E., and Preston, J.} {60} {Proc. Roy. Soc. {\rm
London}}  {257} {250}

%3
\r{Clusius, K., and Loza C. G.} {59} {Z. Naturforsch.} {14A} {23}

%4
\r{Geballe, T. H., Matthias, B. T., Clogston, A. M., Williams, H. J.,
Sherwood, R. C., and Maita, J. P.} {66} {J. Appl. Phys.} {37} {1181}

%5
\r{Geldart, D. J. W., and Taylor, R.} {70} {Canad. J. Phys.} {48} {155}

%6
\r{Gornostyrev, Yu. N., Katsnelson, M. I., Mikhin, A. G., Osetskii, Yu. N.,
and Trefilov, A. V.} {94} {Fizika metall. Metallovede.} {77(2)} {79}

%7
\r{Greenberg, B. A., Katsnelson, M. I., Koreshkov, V. G., Osetskii, Yu. N.,
Peschan\-skikh, G. V., Trefilov, A. V., Shamanaev, Yu. F., and Yakovenkova
L. I.
} {90} {Phys. Stat. Sol. (b)} {158} {441}

%8
\rr {Heine, V., and Weare, D.} {70} {Solid State Physics
{\rm( Acad. Press, N. Y.)}} {24} {}

%9
\rr {Hultgren, R., Decai, R. D., Hawkins, D. T., Gleiser, M., Kelley,
K. K., and Wagman, D. D.} {73} {Selected values of the thermodynamic
properties of elements {\rm(Metals Park, Ohio: Amer. Soc. for Metals)}}
{}{}

%10
\r {Ivanov, A. S., Katsnelson, M. I., Mikhin, A. G., Osetskii, Yu. N.,
Rumyantsev, A. Yu., Trefilov, A. V., Shamanaev, and Yu. F., Yakovenkova, L. I.}
{94} {Phil. Mag.} {B69} {1183}

%11
\rr  {Izymov, Yu. A., and Chernoplekov, N. A.} {95} {Neutron spectroscopy}
{} {(Plenum Press, N. Y.)}

%12
\r {Junod, A.} {80} {Solid State Communications} {33} {55}

%13
\r {Katsnelson, M. I., Naumov, I. I., and Trefilov, A. V.}
 {94} {Phase Transitions}  {B49} {143}

%14
\r {Kraynukov, S. N., Khotkevich, A. N., Yanson, I. K., Zhalko-Titarenko,
A. V., Antonov, V. N., Nemoshkalenko, V. V., Milman, V. Yu., Khlopkin,
M. N., and Shitikov, Yu. L.} {89} {Fizika Tverdogo Tela} {31} {iss. 3, 123}

%15
\r {Maksimov, E. G., and Savrasov, S. Yu.} {95} {Uspekhi Fiz. Nauk}
{165} {773}

%16
\rr {Maradudin, A. A., Montroll, E. W., and Weiss, G. H.} {63}
{Theory of lattice dynamics in the harmonic approximation} {}
{(Acad. Pres. N. Y.)}

%17
\r {Mirmelstein, A. V., Karkin, A. E., Arkhipov, V. E., and Voronin, V. I.}
{84} {Fizika Metallov Metallovede} {58} {1008}

%18
\r{Naumov, V. N.} {94} {Phys. Rev. B} {49} {13247}

%19
\r{Nemoshkalenko, V. V., Milman, V. Yu., Zhalko-Titarenko, Antonov, V. N.,
and Shitikov, Yu. L.} {88} {Pis'ma ZhETF} {47} {245}

%20
\rr{Reznitskii, L. A.} {81} {Calorimetry of Solids
{\rm (Moscow, MSU, in Russian)}} {} {p184}

%21
\r{Vaks, V. G., Kapinos, V. G., Osetskii, Yu. N., Samoliuk, G. D., and
Trefilov, A. V.} {89} {Fizika Tverd. Tela} {31} {139}

%22
\r{Vaks, V. G., Kravchuk, S. P., and Trefilov, A. V.} {80} {J. Phys.} {F10}
{2325}

%23
\r{Vaks, V. G., and Trefilov, A. V.} {77} {Fizika Tverd. Tela} {19} {244}

%24
\r{Vaks, V. G., Zarochentsev, E. V., Kravchuk, S. P., Safronov, V. P.,
and Trefilov, A. V.} {78} {Phys. Stat. Sol. (b)} {85} {63}

%25
\rr{Wolcott, N. M.} {56} {Conference de Physique des Basses Temperatures
{\rm (Institute de Froid, Paris, 1956)}} {} {286}

%26
%\rr{Yanson, I. K., Khotkevich, A. V., Kraynukov, S. N., Nemoshkalenko, V. V.,
%Zhalko-Titarenko, A. V., Milman, V. Yu.} {88} {Physics of Transition
%Metals {\rm (Kiev, Naukova Dumka, in Russian (ed. Baryahtar))}} {}
%{p170}

%26
\r{Zhalko-Titarenko, A. V., Milman, V. Yu., A. V., Antonov, V. N.,
Nemoshkalenko, Khotkevich, A. V., Kraynukov} {89} {Metallofizika} {11}
{iss. 5, 25}

%27
\rr {Zinoviev, V. E.} {89} {Properties of metals at high temperatures} {}
{Moscow, Metallurgia, in Russian}

\newpage
\begin{center}
{\sc Captions to tables of the paper by Katsnelson \ea
``Peculiarities of phonon spectra \dots ''} \end{center} \vskip 5mm

Table 1. Pseudopotential parameters (in atomic units) in relations (2), (4).

Table 2. Values of energy $E$, elastic moduli $B_{ik}$, their derivatives
over pressure, Debye temperature at $T\to 0$ $\Theta_{LT}$, Gruneisen
parameter $\gamma_0$ and maximum phonon frequency $\omega_{max}$
for Ir. $E$ --- Ry/at, P,B
--- 10$^{11}$ dyn/cm$^2$, $\Theta_{LT}$ --- K, $\omega_{max}$ --- THz.
AH corresponds to the values calaulated by the Animalu-Heine
pseudopotential, 4c --- the same with the addition (4c).

Table 3. Values of energy $E$, elastic moduli $B_{ik}$, their derivatives
over pressure, Debye temperature at $T\to 0$ ($\Theta_{LT}$), Gruneisen
parameter $\gamma_0$ and maximum phonon frequency for Rh $\omega_{max}$.
$E$ --- Ry/at, P,B
--- 10$^{11}$ dyn/cm$^2$, $\Theta_{LT}$ --- K, $\omega_{max}$ --- THz.

Table 4a. Parameters characterizing the experimental heat capacity at a
constant pressure, low temperature and high temperature limits of the Debye
temperature and the phonon spectrum momenta  for Ir.
AH corresponds to the values calculated by the Animalu-Heine
pseudopotential, 4c --- the same with the addition (4c).
[Clu] and [Geb] are data sets from Clusius \ea 1959 and Geballe \ea 1966
respectively. Subcolumn ``this paper'' in the column ``experiment''
containes values derived from the experimental data using the technique
described in \S\ 4.

Table 4b. Parameters characterizing the experimental heat capacity at a
constant pressure, low temperature and high temperature limits of the Debye
temperature and the phonon spectrum momenta  for Rh.
AH corresponds to the values calculated by the Animalu-Heine
pseudopotential, 4c --- the same with the addition (4c).
[Clu], [Geb] and [Wol] are data sets from Clusius \ea 1959, Geballe \ea 1966
and Wolcott 1956
respectively. Subcolumn ``this paper'' in the column ``experiment''
containes values derived from the experimental data using the technique
described in \S\ 4.

\newpage \begin{center}
{\sc Figure captions to the paper by Katsnelson \ea
``Peculiarities of phonon spectra \dots ''} \end{center} \vskip 5mm

Fig.1 Section of the Fermi surface of Ir.

Fig.2 Non-uniform susceptability $\chi({\bf q})$ in directions
$\langle 110  \rangle$ (a) and $\langle 110  \rangle$ (b). Numbers in fig. 2b
designate contributions from the corresponding branches.

Fig.3 Phonon spectrum in Ir. Dashed line --- the phonon spectrum calculated
by the Animalu-Heine pseudopotential, solid line --- the same with the
addition (4c). Experimental points ($\diamond$) are taken from Ivanov \ea
1994. Arrows indicate the Kohn singularities in the approximation of free
electrons.

Fig.4 Phonon spectrum in Rh. Dashed line --- the phonon spectrum calculated
by the Animalu-Heine pseudopotential, solid line --- the same with the
addition (4c).
Arrows indicate the Kohn singularities in the approximation of free
electrons.

Fig.5 Dependence of the lattice heat capacity at a constant volume (a) and the
effective Debye temperature (b) on temperature for Ir. Dashed line
corresponds to the values calculated by the Animalu-Heine pseudopotential,
Solid line --- the same with the addition (4c). The experimental values
derived in this paper denoted as $\diamond$. $T_{pl}=\hbar\omega_{pl}=
\hbar(4 \pi Z e^2 /M \Omega_0) ^{1/2}$, $\omega_{pl}$ is the ionic plasma
frequency.

Fig.6 Dependence of the lattice heat capacity at a constant volume (a) and the
effective Debye temperature (b) on temperature for Rh. Dashed line
corresponds to the values calculated by the Animalu-Heine pseudopotential,
Solid line --- the same with the addition (4c). The experimental values
derived in this paper denoted as $\diamond$. $T_{pl}=\hbar\omega_{pl}=
\hbar(4 \pi Z e^2 /M \Omega_0) ^{1/2}$, $\omega_{pl}$ is the ionic plasma
frequency.

Fig.7 Theoretical (---) and experimental ($\diamond$) temperature
dependences of the lattice heat capacity at a constant volume (a) and the
effective Debye temperature (b) for Na from Vaks \ea 1978.
$T_{pl}=\hbar\omega_{pl}=
\hbar(4 \pi Z e^2 /M \Omega_0) ^{1/2}$, $\omega_{pl}$ is the ionic plasma
frequency.

Fig.8 Average square of atomic displacement in Ir (a) and Rh (b).
Dashed line
corresponds to the values calculated by the Animalu-Heine pseudopotential,
Solid line --- the same with the addition (4c).
$T_{pl}=\hbar\omega_{pl}=
\hbar(4 \pi Z e^2 /M \Omega_0) ^{1/2}$, $\omega_{pl}$ is the ionic plasma
frequency.
\newpage
\hoffset -20mm
table 1  
\vskip 5mm
%$\Omega_0$, $r_0$, $\beta$ in atomic units

\begin{tabular} {|c |c | c| c| c| c|c| c| }
\hline
metal & $\Omega_0$ & Z & $\xi$ & $r_0$ & $V_0$& $\alpha$ & $\beta$ \\
\hline
Ir AH & 95.88 & 4.5 & 0.3& 2.696 & -1.243 &  &  \\
\hline
Ir 4c & 95.88 & 4.5 & 0.3& 2.696 & -1.243 & 1.285 & -4.29$\cdot 10^{-3}$ \\
\hline
Rh AH &92.57 & 3.86 & 0.3 & 2.678 & -1.251 & & \\
\hline
Rh 4c &92.57 & 3.86 & 0.3 & 2.678 & -1.251 & 1.276 &-6.76$\cdot 10^{-3}$ \\
\hline
\end{tabular}

\vskip 5mm
table 2 %Calculations results for lattice properties of Ir
\vskip 5mm
%E --- Ry/at; P,B --- $10^{11}$ $dyn/cm ^2$; $\Theta_D$ --- K;
%$\omega_{max}$ --- THz

\begin{tabular} { |c|c|c|c|c|c|c|c|c|c|c|c|c|c|}
\hline
 & E & P & $B_{st} $ & $B_{dyn}$ & $B_{33}$ & $B_{44}$ &
 ${\partial B_{33} \over \partial P}$ &
 ${\partial B_{44} \over \partial P}$ &
 ${\partial B_{st} \over \partial P}$ &
 ${\partial B_{dyn} \over \partial P}$ &
 $ \Theta_D$ & $\gamma_0$ & $\omega_{{max}(X)}$ \\
\hline
AH & -9.064 & 0.00 & 31.27 & 35.17 &14.01 &26.69 &
2.50 & 4.63 & 5.95 & 5.90 & 409 & 1.73 & 8.06\\
\hline
4c & -9.060 & 0.00 & 29.26 & 33.14 & 14.40 & 24.43 &
2.21 & 5.77 & 6.71 & 6.60 & 401 & 1.99 & 6.98\\
\hline
exper & --- &0.00 & 37.1 &37.1 &17.1 & 26.9 &
--- & --- & --- & --- & 425 & --- & 6.91
\\
\hline
\end{tabular}

\vskip 5mm

table 3 %Calculations results for lattice properties of Rh

\vskip 5mm
\begin{tabular} { |c|c|c|c|c|c|c|c|c|c|c|c|c|c|}
\hline
 & E & P & $B_{st} $ & $B_{dyn}$ & $B_{33}$ & $B_{44}$ &
 ${\partial B_{33} \over \partial P}$ &
 ${\partial B_{44} \over \partial P}$ &
 ${\partial B_{st} \over \partial P}$ &
 ${\partial B_{dyn} \over \partial P}$ &
 $ \Theta_D$ & $\gamma_0$ & $\omega_{{max}(X)}$ \\
\hline
AH & -6.860 & 0.00 & 25.33 & 29.02 &10.97 &22.15 &
2.09 & 4.30 & 5.70 & 5.51 & 500 & 1.54 & 10.03\\
\hline
4c & -6.857 & 0.00 & 22.30 & 25.91 & 11.49 & 18.76 &
1.60 & 6.40 & 7.21 & 6.87 & 482 & 2.05 & 7.70\\
\hline
exper & --- &0.00 & 26.9 &26.9 &11.5 & 19.4 &
--- & --- & --- & --- & 482 & --- & 7.63
\\
\hline
\end{tabular}

%\end{document}

\newpage
table 4a  \vskip 5 mm

     \begin{tabular}{|l||c|c|c|c|c|}
     \hline
& \multicolumn {3}{c|} {experiment}& \multicolumn{2} {c|} {theory, this
paper} \\
\cline{2-6}
     parameter &  this paper&[Clu]&[Geb]& AH & 4c \\
     \hline
$ \gamma$, mJ mol$^{-1}$K$^{-2}$& 3.26   & 3.18   &3.20&        &       \\
$ \beta$, mJ mol$^{-1}$K$^{-4}$  & 0.0228 & 0.0244 &    &        &       \\
$ \alpha$, mJ mol$^{-1}$K$^{-6}$ & 0.00005&        &    &        &       \\
$ \Theta_{LT},$ K       & 440    & 430    &    & 409    & 401   \\
$ \Theta_{HT},$ K       & 290    & 289    &    & 355    & 321   \\
$ A,$ mJ mol$^{-1}$K$^{-2}$      & 4.7    &        &    &        &       \\
$ \Omega_{log},$ K      & 192    &        &    & 227    & 214   \\
$ \Omega_{-2},$ K       & 188    &        &    & 222    & 213   \\
$ \Omega_{-1},$ K       & 202    &        &    & 240    & 224   \\
$ \Omega_1,$ K          & 217    &        &    & 260    & 241   \\
$ \Omega_2,$ K          & 225    &        &    & 267    & 248   \\
$ \Omega_4,$ K          & 236    &        &    & 276    & 261   \\
$ \Omega_*$, K          & 240    &        &    &        &       \\
\hline
\end{tabular}

\vskip 5mm
table 4b
\vskip 5mm

     \begin{tabular}{|l||c|c|c|c|c|c|}
     \hline
& \multicolumn {4}{c|} {experiment}& \multicolumn{2} {c|} {theory, this
paper} \\
\cline{2-7}
     parameter &
this paper &[Clu]&[Bud]&[Wol]& AH & 4c \\
     \hline
$ \gamma$, mJ mol$^{-1}$K$^{-2}$ & 4.95   & 4.18  &4.65 &4.90 &      &\\
$ \beta$, mJ mol$^{-1}$K$^{-4}$  & 0.0195 & 0.0213&     &     &      &\\
$ \alpha$, mJ mol$^{-1}$K$^{-6}$ & 0.00002&       &     &     &      &\\
$ \Theta_{LT},$ K       & 505    & 450   & 512 &     & 500   & 482 \\
$ \Theta_{HT},$ K       & 345    & 346   &     &     & 437  &362\\
$ A,$ mJ mol$^{-1}$K$^{-2}$      & 5.5    &       &     &     &      &\\
$ \Omega_{log},$ K      & 227    &       &     &     & 283  &246\\
$ \Omega_{-2},$ K       & 222    &       &     &     & 281  &245\\
$ \Omega_{-1},$ K       & 240    &       &     &     & 299  &256\\
$ \Omega_1,$ K          & 260    &       &     &     & 328  &273\\
$ \Omega_2,$ K          & 267    &       &     &     & 339  &280\\
$ \Omega_4,$ K          & 276    &       &     &     & 359  &293\\
$ \Omega_*$, K          & 278    &       &     &     &      &\\
\hline
\end{tabular}

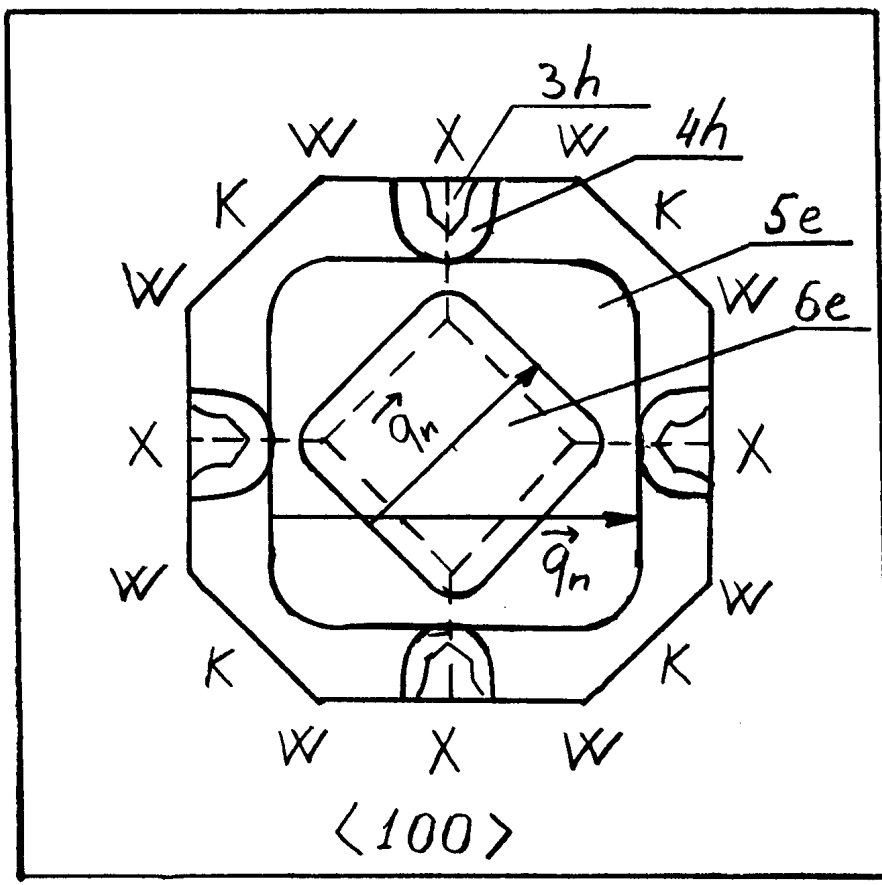
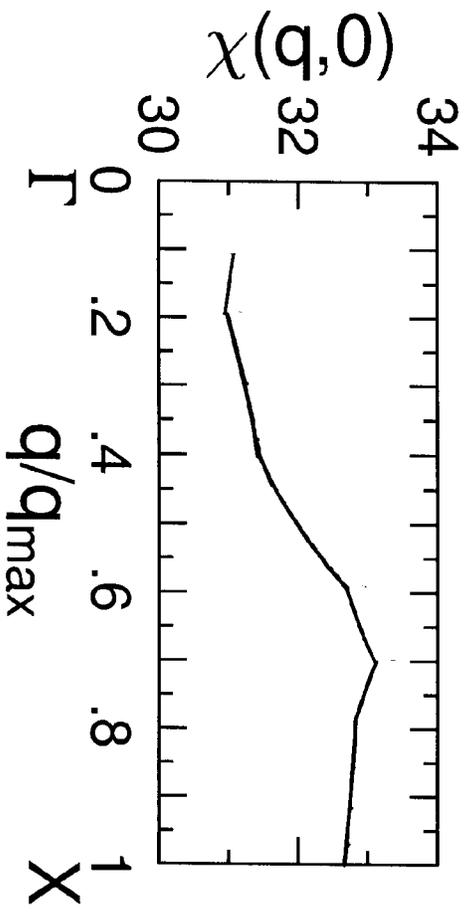
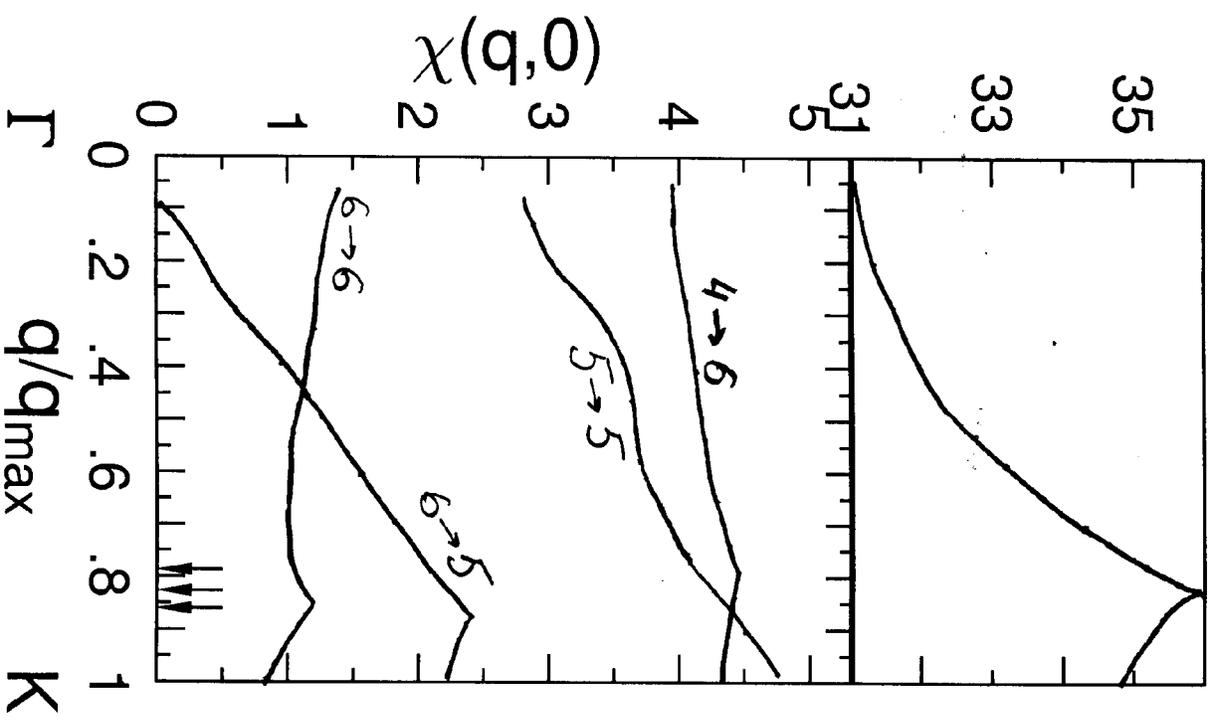
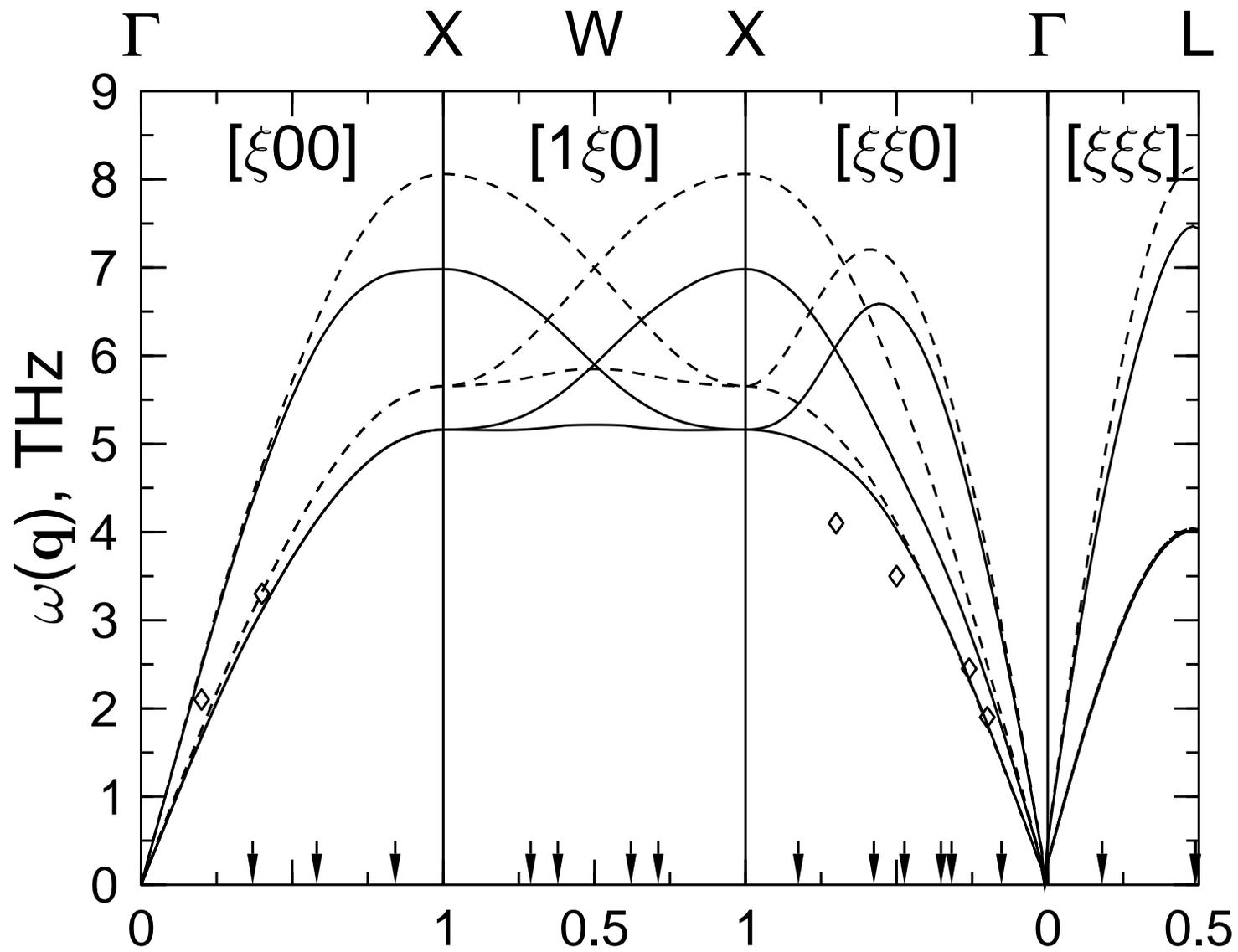
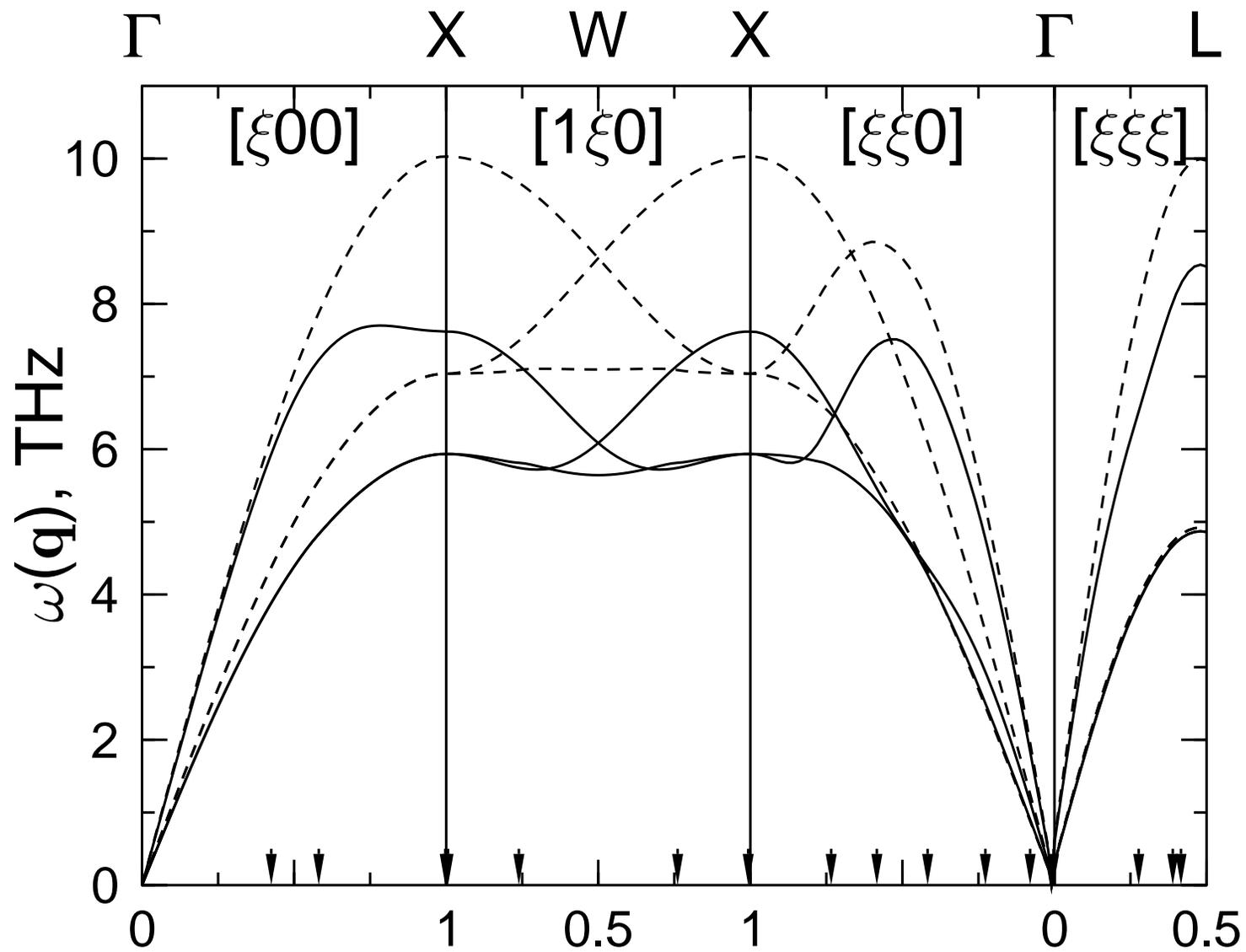
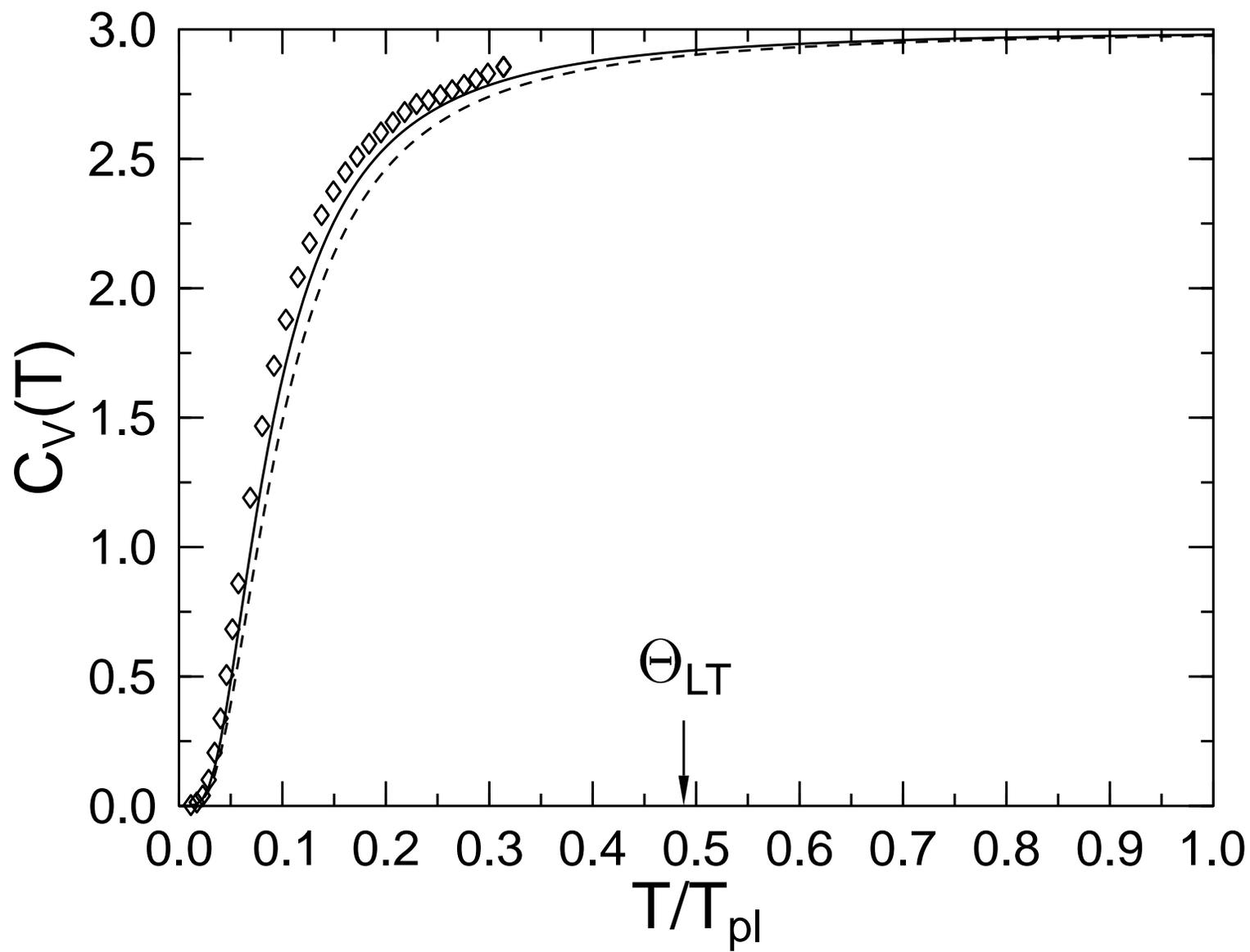
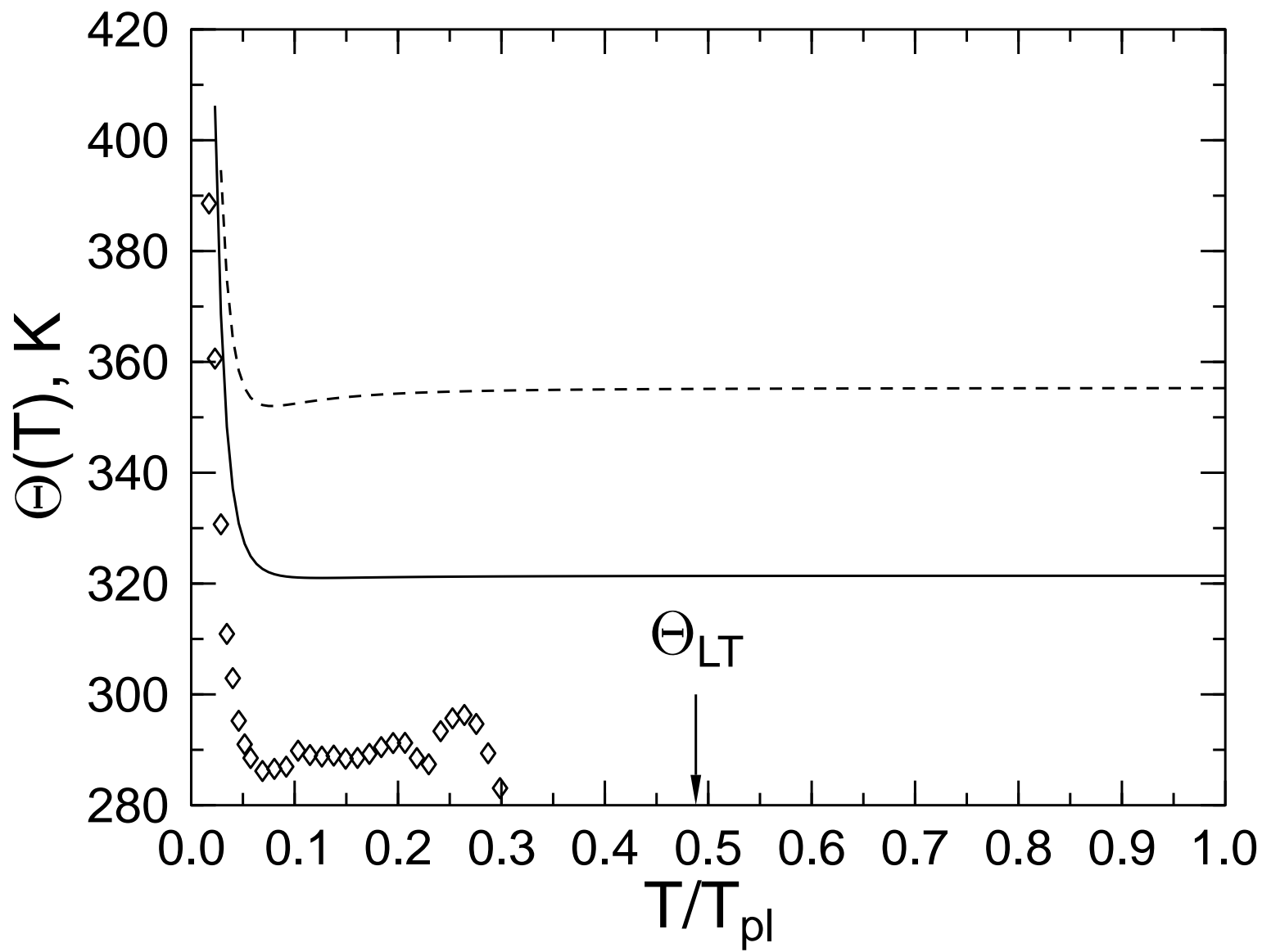
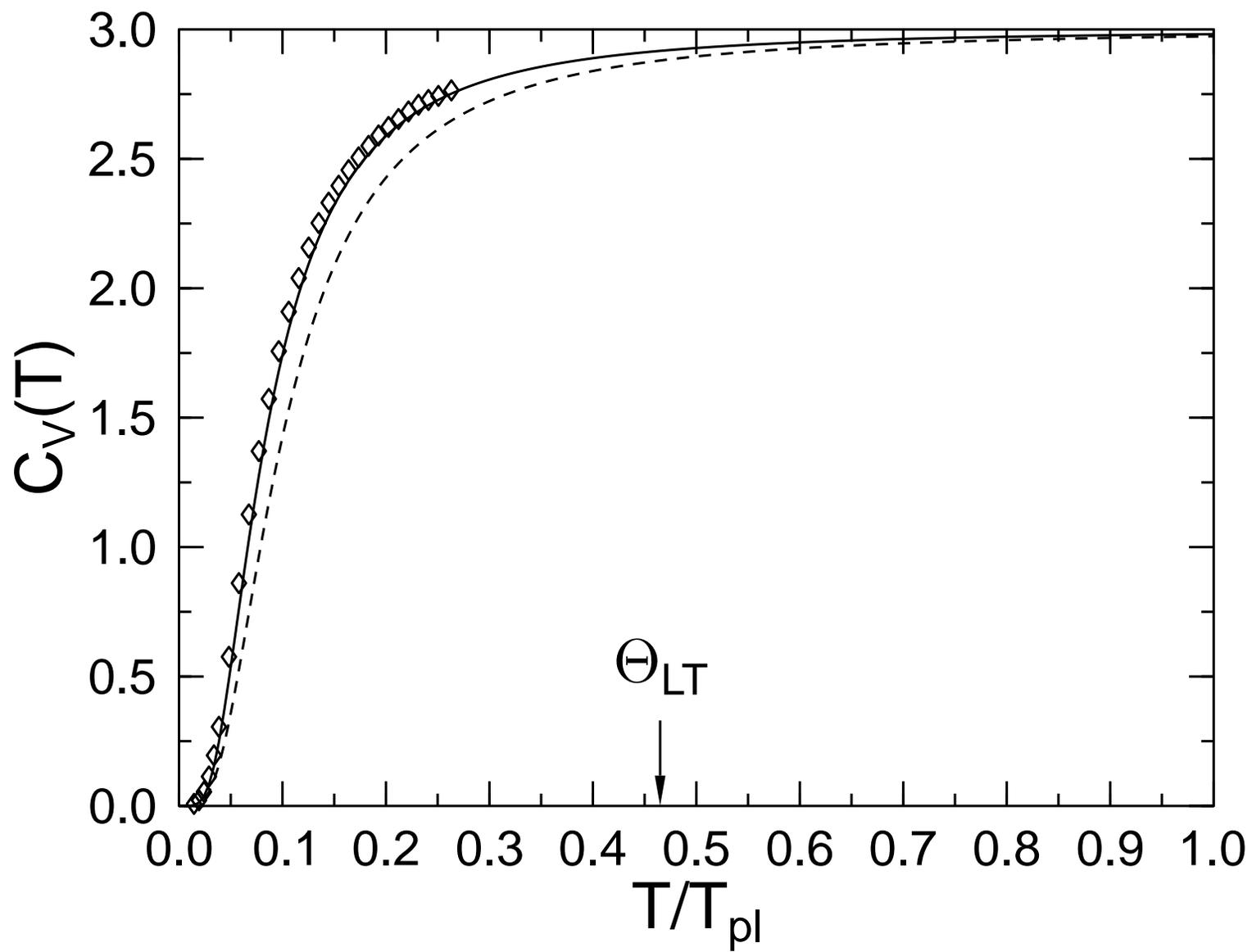
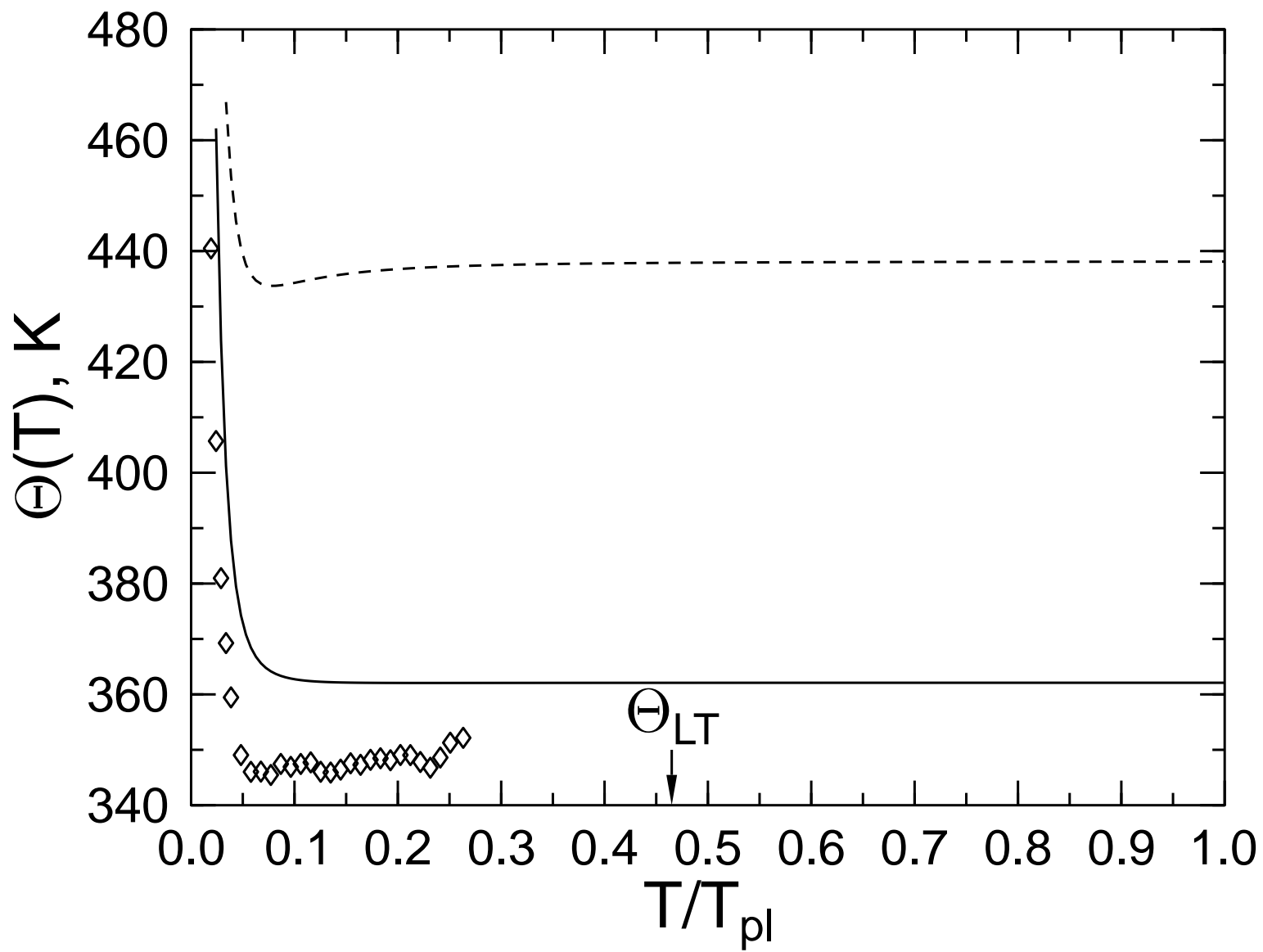
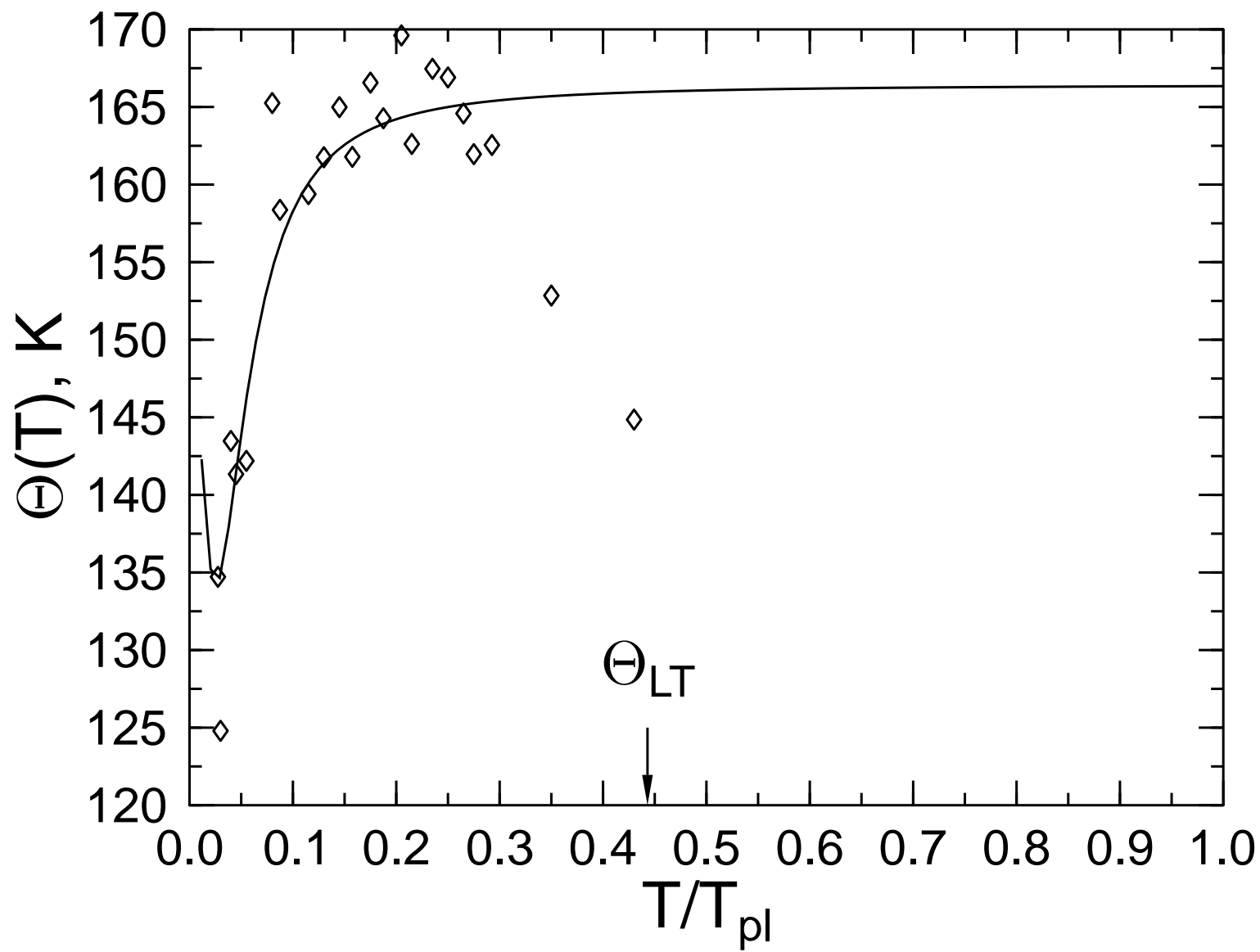
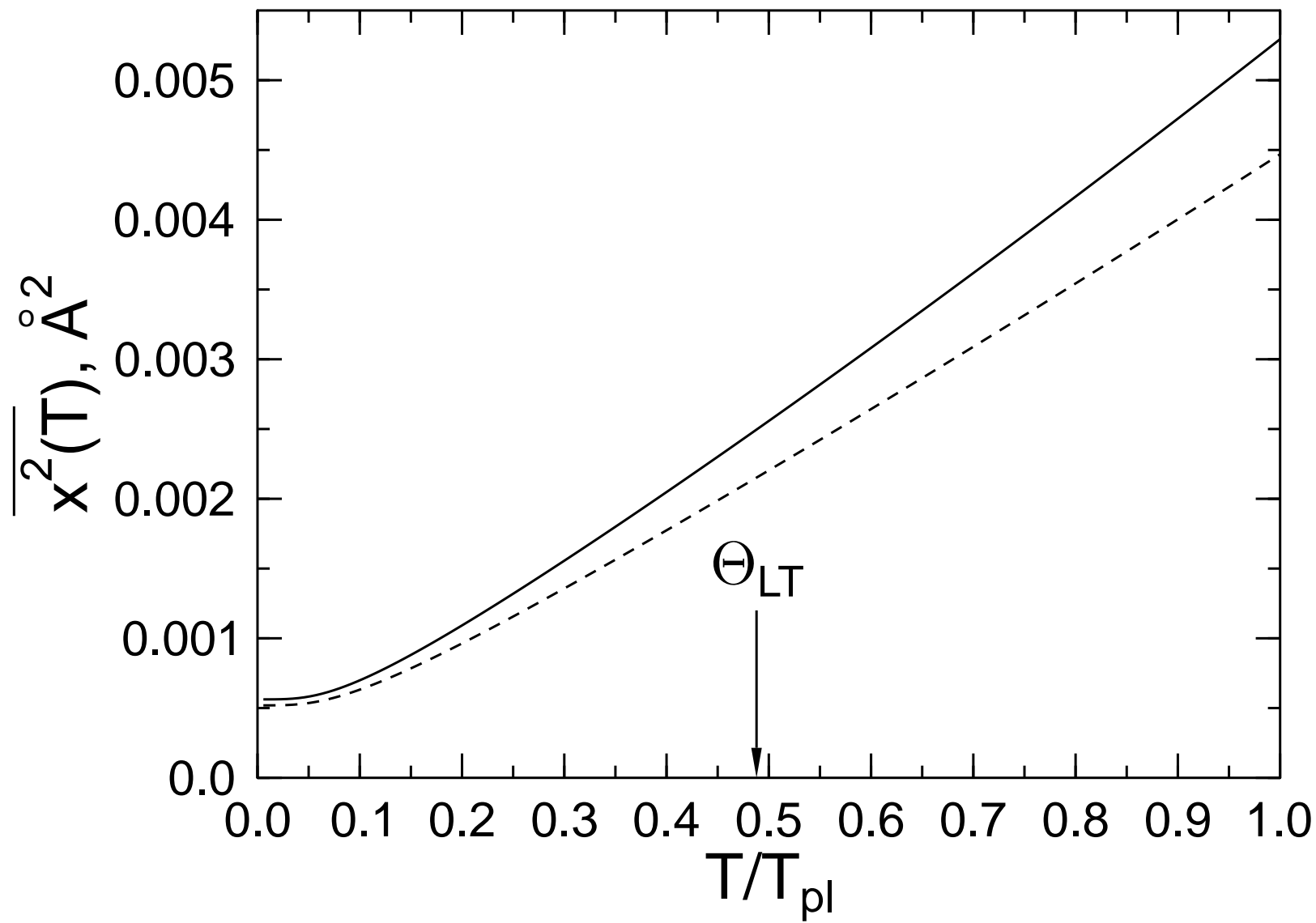
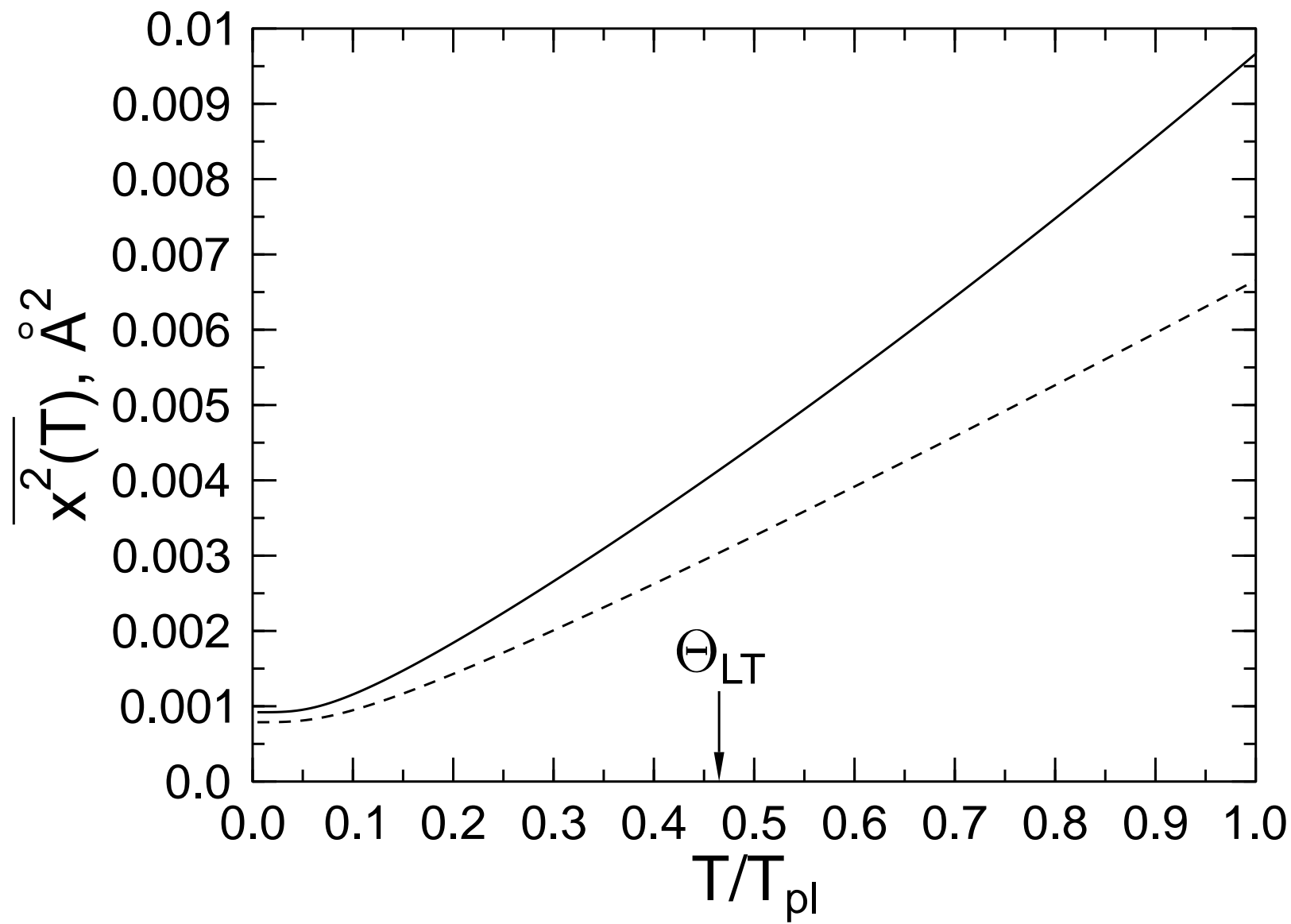
\end{document}